\documentclass[aps,prstab,twocolumn,groupedaddress,amsmath,nobibnotes,nofootinbib]{revtex4-1}
\usepackage{graphicx}
\usepackage{float}
\usepackage{subcaption}
\usepackage{caption}
\usepackage{tabularx}
\usepackage{threeparttablex}
\usepackage{natbib}
\newcommand\Tstrut{\rule{0pt}{2.6ex}}         % = `top' strut
\newcommand\Bstrut{\rule[-0.9ex]{0pt}{0pt}}   % = `bottom' strut

\begin{document}

% Use the \preprint command to place your local institutional report
% number in the upper righthand corner of the title page in preprint mode.
% Multiple \preprint commands are allowed.
% Use the 'preprintnumbers' class option to override journal defaults
% to display numbers if necessary
%\preprint{}

%Title of paper
%\title{Measurement of the longitudinal emittance of a high-intensity ion beam output by RFQ}
\title{High dimensional characterization of the longitudinal phase space formed in a radio frequency quadrupole}

% repeat the \author .. \affiliation  etc. as needed
% \email, \thanks, \homepage, \altaffiliation all apply to the current
% author. Explanatory text should go in the []'s, actual e-mail
% address or url should go in the {}'s for \email and \homepage.
% Please use the appropriate macro foreach each type of information

% \affiliation command applies to all authors since the last
% \affiliation command. The \affiliation command should follow the
% other information
% \affiliation can be followed by \email, \homepage, \thanks as well.
\author{K. Ruisard\textsuperscript{1}\email{ruisardkj@ornl.gov}, A. Aleksandrov, S. Cousineau, A. Shishlo, V. Tzoganis, A. Zhukov}
\email[]{ruisardkj@ornl.gov}
%\homepage[]{Your web page}
%\thanks{}
\affiliation{\textsuperscript{1}Oak Ridge National Laboratory, Oak Ridge, Tennessee 37830, USA}

\date{\today}

%%%%%%%%%%%%%%%%%%%%%%%%%%%%%%%%%%%%%%%%%%%%%%%%%%%%%%%%%%%%%%%%%%%%%%%%%%%%%%%%%%
\begin{abstract}
Modern accelerator front ends almost exclusively include radio-frequency quadrupoles for initial capture and focusing of low-energy CW beams. 
Dynamics in the RFQ define the longitudinal bunch parameters.
Simulation of the SNS RFQ with PARMTEQ seeded with a realistic LEBT distribution produces a 2.5 MeV, 40 mA H- beam with root-mean-square emittance of 130 deg-keV.
A detailed characterization of the longitudinal phase space is made, including a novel study of the dependence of longitudinal emittance on transverse coordinates.
This work introduces a new \textit{virtual slit} technique that provides sub-slit resolution in an energy spectrometer as well as an approach for visualizing 4D phase space data.
Through simulation and measurement, the RFQ-formed bunch is confirmed to have significant internal correlated structure. 
The high-dimensional features are shown to be in qualitative agreement. 
However, the measured rms emittances are up to 30\% lower than predicted, closer to the design value of 95 deg-keV. 
\end{abstract}
%%%%%%%%%%%%%%%%%%%%%%%%%%%%%%%%%%%%%%%%%%%%%%%%%%%%%%%%%%%%%%%%%%%%%%%%%%%%%%%%%%

\maketitle

%#############################################################################
\section{Introduction}

Low-level beam loss is a fact of life in high-intensity accelerator facilities. 
Controlling and reducing losses to maintain a safe accelerator environment is achieved mainly through online empirical optimization. 
One tool currently missing from the arsenal is high fidelity simulation capable of predicting these losses.
As high-intensity accelerators continue towards higher demands in beam power, the need for this capability becomes more accute \cite{Henderson2010}. 
A strong contributor for losses in a linear accelerator is beam halo \cite{Fedotov2003b, Gluckstern1994}. 
As the beam distribution is both the source and driver of halo particles, loss-level simulation accuracy will require an equally accurate representation of the initial distribution.

There are two approaches for generating a realistic front-end initial distribution.
One is a pure ``end-to-end" approach, which applies self-consistent simulation of the entire beam transport system starting at or downstream of the ion source. 
This may include self-consistent modeling of the ion source/extraction electrodes, abstracting to an idealized distribution in the Low Energy Beam Transport (LEBT) section, or measuring the transverse phase space of the LEBT beam.
The LEBT distribution is then propagated through the radio frequency quadrupole (RFQ). 
The longitudinal bunch is formed inside the RFQ, where dynamics are complicated by nonlinear focusing from both the vane structure and space charge. 
The complexity of the simulation may limit the accuracy of output bunch, as there is large potential for errors.

Previous work at Los Alamos \cite{Allen2002,Qiang2002}, found that the bunch generated through simulation of the RFQ was not sufficiently accurate to model beam dynamics in a medium-energy transport line (MEBT). 
Particularly, transport with mismatched optics was seen to be very sensitive to the initial distribution. 
In this effort, correcting the simulated bunch by rescaling to match observed rms parameters was not sufficient to reach good agreement. 

%\vfill\eject % column break
Alternatively, a bunch may be generated from measurements in the MEBT, after the longitudinal bunch is fully formed but at an energy where detailed measurements are still possible. 
Characterization of the beam in the MEBT circumvents the need to model the complex internal RFQ dynamics, and arguably results in more a representative bunch. 
However, internal correlations are neglected in this approach, which typically relies on 2D projections or Twiss parameters (for example, see \cite{Groening2008,Roy2018c}).
Direct measurement of the 6D beam distribution has been demonstrated \cite{Cathey2018}, but for now end-to-end simulation remains the most accessible option for generating fully-correlated particle coordinates.

Knowledge of the fully-correlated distribution is necessary to accurately portray a bunched beam.
As this article will demonstrate, the bunch formed in the RFQ has non-trivial internal structure.
The formation of the longitudinal phase space is mediated by the space charge force, which couples the three planes \cite{Cathey2018,Wang1982,Fedotov1999a,Chao2003}.
Observations reported here show how both the bunch shape and energy profile to vary with distance from the high-density core. 
As core mismatch is known to excite halo growth \cite{Gluckstern1994,Wangler1998}, it is almost certain that loss-predictive simulations will require knowledge of the realistic 6D structure.

Given that simulation is the most readily available source of fully correlated bunches, one may wonder to what extent simulation reproduces the true 6D structure.
If there is discrepancy in the rms predictions, can one still trust the high-dimensional features?
To begin addressing this question, a detailed characterization of longitudinal phase space is compared with predictions from RFQ simulation.
While the primary metric is the rms emittance, it is applied to slice emittances rather than full emittance.
By varying slice dimensionality and location, the dependence of longitudinal emittance on transverse coordinates is studied.
This provides a method to visualize the high-dimensional features inside the RFQ bunch.

\clearpage
\subsection{SNS Beam Test Facility}

The SNS Beam Test Facility is a one-to-one replica of the SNS front-end, composed of 50 mA H- ion source, 65 kV LEBT, 402.5 MHz RFQ and 1.3 meters of MEBT quadrupoles. 
In addition, the BTF is equipped with extensive diagnostics enabling direct measurement of the 6D phase space distribution.
This phase space diagnostic includes two pairs of vertical/horizontal slits for isolation of the transverse phase space coordinates, followed by an energy spectrometer comprised of a $90^{\circ}$ dipole and vertical slit. 
Finally, time-of-arrival measurement of secondary electron emission from a beam-intersecting wire serves as a bunch shape monitor. 
The energy spectrometer and bunch shape monitor are used for longitudinal phase space measurements described here. 

Accelerator physics studies at the BTF are motivated by the goal of demonstrating halo-predictive simulation. 
Ongoing efforts have followed a three-pronged approach: extensive characterization of the initial MEBT beam distribution \cite{Cathey2018}, deployment of high dynamic range phase space diagnostics for halo detection, and extension of the BTF MEBT to support studies of halo evolution \cite{Zhang2019}.
The work described here falls in the first category, as it addresses the applicability of RFQ simulations to high-fidelity simulation.

\subsection{Emittance Convention}

% [emittance definitions]
With the BTF apparatus for longitudinal emittance measurement, a dynamic range in excess of $10^3$ has been demonstrated. 
As the rms parameters can depend heavily on the threshold, it is necessary to speficy the applied threshold.
In order to standardize emittance values between simulation and measurement, we adopt the following metrics for reporting emittances: 
\begin{itemize}
\item 0.1\% emittance, near dynamic-range limit of measurements, with a threshold applied at 0.1\% of peak density,
\item 1\% emittance, and
\item 10\% emittance, representing the core of the beam.
\end{itemize}

\noindent In simulation it is common to report emittances based on percentage of enclosed particles, such as 90\%, 99\% emittances. This is distinct from the definitions here. However, the term ``100\% emittance" is still used to indicate inclusion of all simulation particles when no threshold is applied.  

% [phase convention]
In plots of the longitudinal phase space, positive phase corresponds to positive time.
The tail of the bunch, which arrives at a later time than the head, has $\phi_{tail} > \phi_{head}$. 
This phase convention matches the convention used in RFQ simulation.

\subsection{Dimensionality}
% [describe notation for partial projection]

\begin{figure}
\includegraphics[width = 0.36\textwidth]{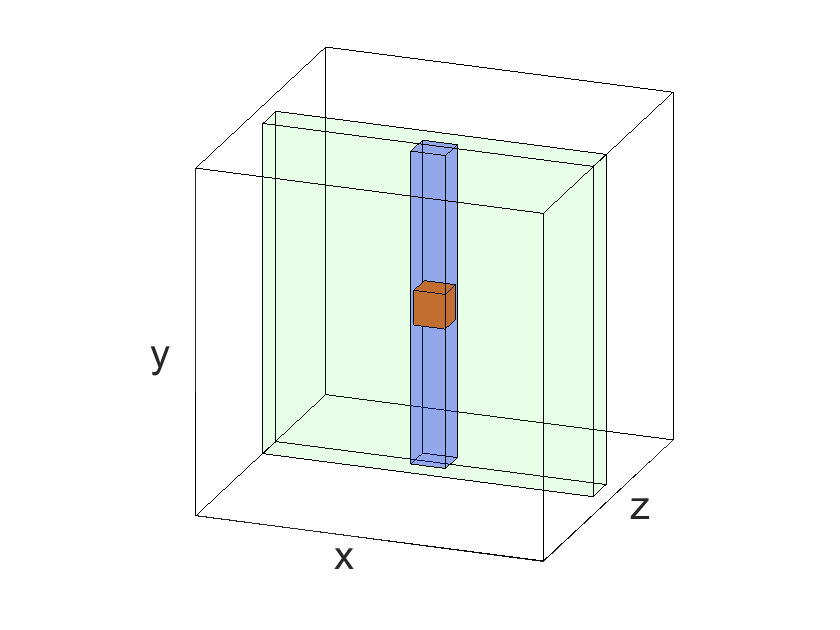}
\caption{\label{fig:cube} Illustration of high dimensional slices in 3D space. A slice of the cube can be made along one dimension (eg, $\tilde{z}=0$, the green volume), two dimensions ($\tilde{z},\tilde{x}=0$, blue) or all three dimensions ($\tilde{z},\tilde{x},\tilde{y}=0$, red).
If the 1D and 2D slice volumes are projected onto the $y,z$ plane, they will be indistiguishable, as the slice $\tilde{x}$ is along a hidden (projected) dimension.}
\end{figure}

In addition to threshold, the reported emittances will depend on the \textit{dimensionality} of the phase space used for the calculation. 
Typically, emittances are calculated for the \textit{fully projected} phase space. That is, the $\left( \phi,w \right)$ coordinates of every particle regardless of location in transverse phase space are included.
The BTF longitudinal emittance apparatus is set up to sample phase space for a \textit{slice} in the transverse coordinates. 
This is referred to as a \textit{partially projected} phase space, and the resulting emittance a partial or slice emittance.
It should be immediately clear that many unique partial projections of the longitudinal phase space are possible. 
Slices can be made in one or several or all of the transverse coordinates, and the slices can be taken at varying distance from the beam core. 
Figure \ref{fig:cube} illustrates the slice concept in three dimensions.
As slices are generally made on hidden (unplotted) dimensions, it is not explicitly apparent whether a phase space plot represents a full or n-dimensional partial projection.

Therefore, when discussing slice emittances, it is important to indicate both the dimensionality (how many slices) and the slice location. 
Here, $f$ indicates the full 6D phase space density $f(x,x',y,y',\phi,w)$.
$\hat f$ indicates a partial projection, where a slice is made in at least one dimension.
A tilde is used to indicate a slice of finite width. 
Unless otherwise specified, the slice width is equal to the width determined by physical apertures in the measurement.
If coordinates do not appear as arguments or slices of $\hat f$, the density along that coordinate is integrated.
With this notation, a partially projected longitudinal phase space representing only the particles within a finite slice centered at $x=0$, $x'=0$ is described as $\hat f(\phi,w)|_{\tilde{x},\tilde{x}' = 0}$.

\begin{table}[th]
\caption{rms parameters of the realistic LEBT distribution at the entrance to the RFQ.} \label{tab:lebt}
\begin{ruledtabular}
\begin{tabular}{l lll@{\extracolsep{\fill}} }  
Quantity & horizontal & vertical &\Tstrut \\ \hline
$\epsilon$ [norm, mm-mrad] &  0.24 & 0.24  &	\Tstrut \\  
$\alpha$	& 1.08 & 0.97 &\\
$\beta$  [mm/rad] 	& 51 & 49 &\Bstrut \\
\end{tabular}
\end{ruledtabular}
\end{table}
\begin{table}[t]
\caption{rms parameters of the bunch at the RFQ output, based on Parmteq simulation for different input distributions.
Transmission calculated for 50 mA input current. Longitudinal emittances are 100\%, unnorm. }
\label{tab:parmteq}
\begin{ruledtabular}
\begin{tabular}{l @{\extracolsep{\fill}}cccc} %\Tstrut\Bstrut\\ 
Input distribution				&  Realistic  & 4D Waterbag & KV & \Tstrut \\ \hline
Transmission					& 82\%	& 90\%	& 88\% & \Tstrut \\
$\epsilon_z$ [deg-keV] 		& 127.1 & 88.8 & 102.2 & \Tstrut \\
$\alpha_z$ 					&  0.18 & 0.27 & 0.21 &\\
$\beta_z$  [deg/keV]			& 0.88 & 1.38 & 1.16&\\
$\epsilon_x, \epsilon_y$ [norm, mm-mrad]		& 0.22 & 0.12 &  0.15 & \Tstrut \\
%$\alpha_x$					& 2.12 &  2.78  & 2.45  & \\
%$\beta_x$ [m/rad]				& 0.20 &   0.26 &  0.23 & \\
%$\epsilon_y$  [norm, mm-mrad] 		&  0.22 &  0.12 &  0.15	 & \\
%$\alpha_y$					& 1.64 &  2.15 &  1.93	& \\
%$\beta_y$ [m/rad]				& 0.15 &  0.19 &  0.17  	& \\
\end{tabular}
\end{ruledtabular}
\end{table}

\begin{figure*}[ht]
\begin{minipage}{\textwidth}
\centering
\begin{subfigure}[t]{0.3\textwidth}
	\centering
	\includegraphics[width=\textwidth]{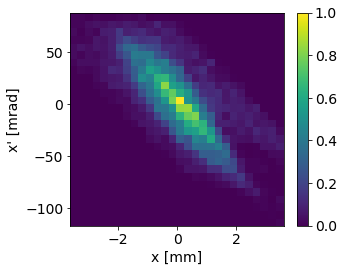}
	\caption{\label{fig:real-rfq-xxp} Horizontal phase space of RFQ input distribution.  }
\end{subfigure}
\begin{subfigure}[t]{0.3\textwidth}
	\centering
	\includegraphics[width=\textwidth]{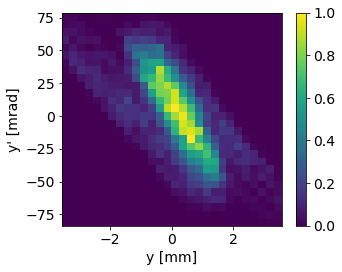}
	\caption{\label{fig:real-rfq-yyp} Vertical phase space of RFQ input distribution.  }
\end{subfigure}
\begin{subfigure}[t]{0.3\textwidth}
	\centering
	\includegraphics[width=0.95\textwidth]{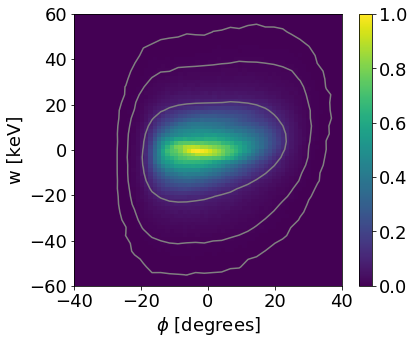}
	\caption{\label{fig:real-rfq-phiw} Fully projected longitudinal phase space at the RFQ exit. }
\end{subfigure}
\caption{\label{fig:real-distr} Fully projected phase space plots for initial (\ref{fig:real-rfq-xxp},\ref{fig:real-rfq-yyp}) and output (\ref{fig:real-rfq-phiw}) distributions from Parmteq simulation. Density is plotted in linear scale. Contours on (\ref{fig:real-rfq-phiw}) show the 10\%, 1\% and 0.1\% threshold levels.}
\end{minipage} 
\end{figure*}

\vfill\eject % column break
%\subsection{Organization of article}
The organization of this article is as follows.
Section \ref{sec:simulation} defines the ``expected distribution," as determined through RFQ simulation. 
After that, the measurement technique is introduced in Section \ref{sec:apparatus}, including accouning for the dominant error sources. 
The largest error is shown to be through point spread in the phase coordinate. Correction of the point spread error is justified through simulation and application of a virtual slit measurement. 
The high dimensional characterization of longitudinal phase space is reported in Section \ref{sec:measurements}. The results show the dependence of the longitudinal slice distribution on RFQ amplitude and transverse coordinates. Finally, Section \ref{sec:comparison} summarizes the comparison between the expected and measured distributions.

%#############################################################################
\section{Simulation} \label{sec:simulation}

\subsection{RFQ output distribution}\label{sec:parmteq}

Original design studies for the SNS RFQ used the Los Alamos code PARMTEQ\cite{Crandall1988a}.
The RFQ accelerates $H^-$ from 65 kV to 2.5 MeV, achieved with vane voltage 83 kV and 449 cells. 
Additionally, the design is constrained to produce $\leq95$ keV-deg at maximum current output.
This goal was met with normalized input emittance $0.2$ mm-mrad and simulation transmission $>90\%$ \cite{Henderson2014,Ratti2000,Ratti2002b}.
In this paper, the PARMTEQ simulation is repeated with an input beam based on LEBT measurements.

The PARMTEQ space charge calculation uses the SCHEFF module with a cylindrical geometry.
Saturation of the PARMTEQ simulation was judged by the rms Twiss parameters of the output bunch. 
40,000 particles and a grid spacing of 10 radial segments and 20 longitudinal segments was sufficient. 
However, for the results reported here up to 5,000,000 macroparticles are used.
The higher particle number was necessary for good statistics when calculating rms emittances for high-dimensional slices.

The input beam is initially mono-energetic with $w\equiv T-T_0=0$ for all particles, and initial uniform random phase.
The transverse distribution is generated from measurements of the horizontal and vertical phase space distributions in the LEBT. 
These measurements were acquired in 2012 for ion source output of 50 mA. The measured distribution is transformed from the measurement plane to the RFQ entrance using a matrix equations. The resulting transverse phase space distributions are plotted in Figures \ref{fig:real-rfq-xxp} and \ref{fig:real-rfq-yyp}. 
The rms Twiss parameters are reported in Table \ref{tab:lebt}.

As the motivation of this study is to address the role of RFQ simulations in high-fidelity modeling, the measurement-based LEBT distribution is used to generate the \textit{expected} distribution, under the assumption that this is the most likely to resemble the actual beam parameters. However, comparison to equivalent transverse waterbag and KV distributions give an indication of possible spread in values due to uncertainty of initial distribution. Equivalent is defined as having identical rms Twiss parameters.

Table \ref{tab:parmteq} compares the rms Twiss parameters for the realistic LEBT distribution against equivalent 4D waterbag and KV distributions. 
For the realistic input distribution, the 100\% rms emittance is $\epsilon_z = 127.1$ deg-keV. 
The rms widths of the fully projected distribution are $\phi_{rms} = 10.6^{\circ}$ and $w_{rms} = 12.2$ keV .
The longitudinal distribution at RFQ exit for the realistic LEBT distribution is shown in Figure \ref{fig:real-rfq-phiw}. 
From this figure, it is apparent that the longitudinal phase space is very far from Gaussian and likely has complex internal structure as well.

\begin{figure}
\centering
\includegraphics[width=0.45\textwidth]{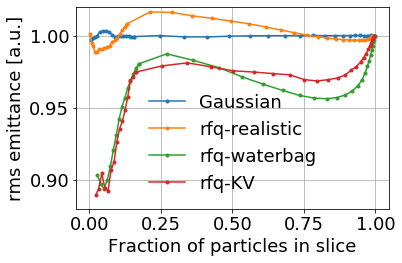}
\caption{\label{fig:slice-width} Dependence of calculated longitudinal emittance on width of the slice in coordinates $\left(x,x'\right)$ centered at $\left(0,0\right)$. 
Width is expressed as fraction of particles that fall within the slice.  Curves are normalized to the right-most point, which correspond to $100\%$ emittances reported in Table \ref{tab:parmteq}.
A correlation-free 6D Gaussian is included to illustrate the effect of particle noise in the narrowest slices. }
\end{figure}

Figure \ref{fig:slice-width} illustrates the relationship between slice and full emittance by plotting the dependence of the rms longitudinal emittance on the width of a transverse $\left( x,x' \right)$ slice centered over the beam core.
The narrowest slices (left-most points) have widths comparable to the measurement resolution, $\tilde{x}=\pm 0.1$mm and $\tilde{x}'=\pm 0.2$mrad.
The output bunch formed for three different intial distributions have very complex structure compared to a Gaussian beam, which would appear as a straight horizontal line. In the waterbag and KV case, the emittance of the core slice is 10\% lower than the full emittance. 
In the realistic case, for which the initial bunch resembles a Gaussian with heavy, nonlinear tails, the core slice is very similar to the full emittance but there are still distinct features. 
For all distributions, the emittance of an arbitrary slice should not be assumed to be representative of the full emittance.

\begin{figure}
\includegraphics[width=0.4\textwidth]{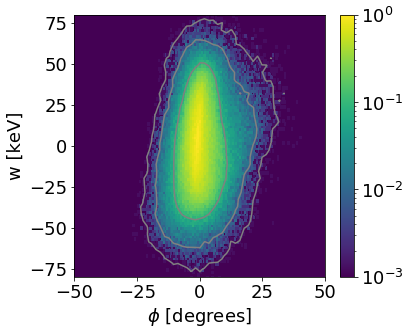}
\caption{\label{fig:expected} 
Expected longitudinal phase space generated via PARMTEQ and PyORBIT propagation of the reaslistic LEBT distribution. Emittance is plotted at the location of the first slit in the MEBT.}
\end{figure}

% expected distribution
\subsection{Expected distribution in the MEBT} \label{sec:expected}
The BTF measurements are made with respect to a plane 1.36 meters downstream of the RFQ. The reference point is the location of the first vertical slit used in the phase space measurement. After this point, at least 98\% of the beam is intercepted. In the remaining 2\% ``beamlet," there should be no contribution from space charge on the beam evolution, and downstream measurements can easily be mapped to this plane via matrix equations. 
For the purposes of comparison, it is considerably more straighforward to propagate the self-consistent 6D Parmteq distribution to the measurement plane than back-propagate the measured phase space. 
With this in mind, the expected distribution is defined as the output from Parmteq simulation seeded with the initial measured LEBT distribution at the plane of the first slit in the emittance apparatus.

\begin{table}[bt!]
\centering
\caption{\label{tab:4quads} Parameters used for four-quadrupole MEBT transport line between RFQ output and plane of first vertical slit.}
\begin{ruledtabular}
\begin{tabular}{l  llll}
Position [m] & $L_{eff}$[m] & $\int B \cdot dl$ [T] & Polarity &\Tstrut\Bstrut\\
\hline  
0.1306 & 0.061 & 1.12 &F&\Tstrut \\
0.3139 & 0.066 & -1.25 &D&\\
0.5751 & 0.096 & 1.08 &F&\\ 
0.7709 & 0.096 & -0.61 &D&\\
\end{tabular}
\end{ruledtabular}
\end{table}

%\vfill\eject % column break
Modeling of the MEBT is done with the particle-in-cell code PyORBIT \cite{Shishlo2015}. 
Between the RFQ and the first slit, the MEBT contains four quadrupoles. A hard-edged model was used, with parameters listed in Table \ref{tab:4quads}.
A stepsize of 1 cm is used for the space charge calculation.

The expected distribution at the measurement plane is plotted in Figure \ref{fig:expected}.
For the fully projected phase space, the 100\% rms emittance is $\epsilon_z = 131$ deg-keV and the rms widths are $5.6^{\circ}$ and $24$ keV. 
The phase width  is reported for the ``shear-corrected" frame, where the linear phase correlation has been subtracted. 
For comparison, the uncorrected phase width at this location is $43^{\circ}$.
The bunch has significantly different aspect ratio than at the RFQ exit ($10.6^{\circ}$ and $12.2$ keV) due to space charge driven debunching in the first 20 cm of transport \cite{Ruisard2020}.  
This causes the energy spread to increase by $\sim 2 \times$ and the phase width to decrease by $\sim 2 \times$.

Table \ref{tab:sim-phi-w} compares the 1\%-thresholded emittance and rms widths for slices of different dimensionality in the expected distribution.
The rms emittance does not have a strong dependence on the dimensionality of the slice. 
In the context of measurement, the variation in emittance values is comparable to the uncertainty in measurement, which will be shown to be around $10 - 15\%$. 
As such, the measured values of core-slice emittances are expected to be very near the fully projected emittance. 
Also apparent in Table \ref{tab:sim-phi-w} is that the energy/phase aspect ratio increases with dimensionality.
The space charge driven debunching is amplified in the high-density core, which is preferentially included in centered, high-dimensional slices.
For this reason, the rms energy spread becomes larger and the rms phase spread smaller for core slices when compared to the fully projected phase space. 

\begin{figure}
\includegraphics[width=0.5\textwidth]{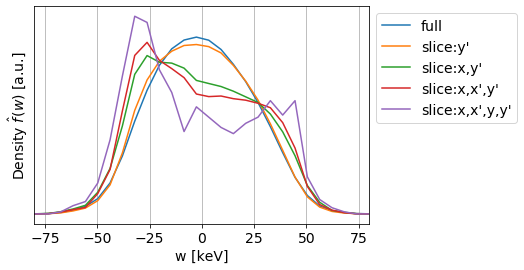}
\caption{\label{fig:expected-energy-profile} 
Dependence of partially projected energy profile $\hat f(w)$ on dimensionality of slice.}
\end{figure}

Figure \ref{fig:expected-energy-profile} shows the dependence of the expected energy profile $\hat f(w) = \int d\phi \hat f(\phi,w)|_{\vec x}$ on the dimensionality of a slice in transverse dimensions.
This illustrates the increase of rms energy width reported in Table \ref{tab:sim-phi-w} as well as the presence of very non-Gaussian internal structure.
This structure strongly resembles with the initial observation of high-dimensional correlations reported in \cite{Cathey2018}.

%
%\begin{figure}
%\begin{minipage}[c]{0.5\textwidth}
\begin{table}[t!]
\centering
\begin{ruledtabular}
\begin{tabular}{llll}
slice & $\epsilon_z$ [deg-keV]& rms $\phi$ [deg] & rms $w$ [keV] \Tstrut\Bstrut\\
\hline  
none (full)						& 122  & 5.4 & 22.8\Tstrut \\ 
$\tilde{y}=0$ 					& 117  & 5.1 & 22.9\\ 
$\tilde{y},\tilde{x}=0$  				& 135 & 5.2 & 25.8 \\  
$\tilde{y},\tilde{x},\tilde{x}'=0$ 			& 128 & 4.9 & 26.4 \\ 
$\tilde{y},\tilde{x},\tilde{x}',\tilde{y}=0$ 	& 144  & 4.9 &29.8 \\  
\end{tabular}
\end{ruledtabular}
\caption{\label{tab:sim-phi-w} Dependence of rms quantities of expected distribution on dimensionality of phase space slice.
All slices are centered over the beam core.
rms values are calculated with 1\% threshold applied. 
For these results, the RFQ simulation is seeded with 5,000,000 particles to obtain good statistics in high-dimensional slices.
The slice widths are twice as large as in measurement for the same reason.}
\end{table}%

%#############################################################################
\section{Apparatus} \label{sec:apparatus}

%\subsection{Longitudinal emittance diagnostic}

\begin{figure*}[th]
\includegraphics[width=0.75 \textwidth]{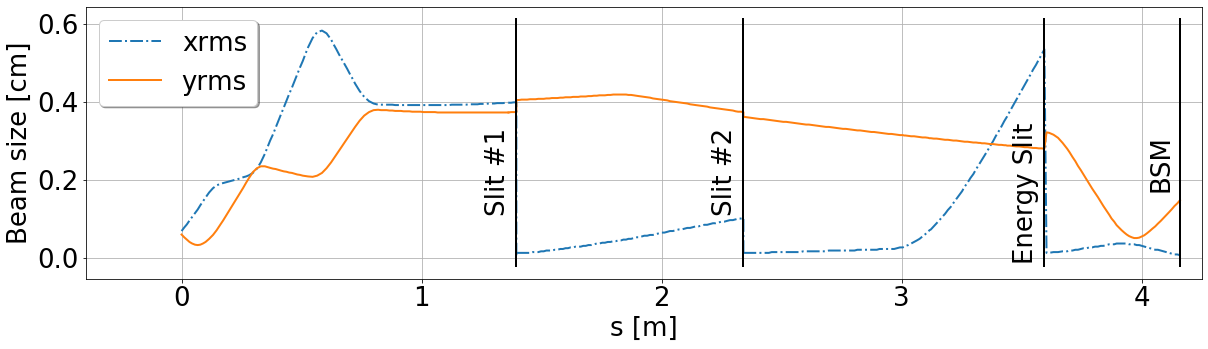}
\caption{\label{fig:optics} Optics view of longitudinal emittance measurement, showing the location and effect of the three vertical slits and bunch shape monitor (BSM). Position $s$ is measured from the exit face of the RFQ. After Slit \#1, the scale of the x-rms curve is mm, rather than cm.  }
\end{figure*}

The apparatus for longitudinal emittance measurement at the SNS BTF is a combination of an energy spectrometer and bunch shape monitor. This apparatus employs a dipole-slit system to isolate a narrow band of energy, followed by a bunch shape monitor to measure the phase distribution. The apparatus was first described in \cite{Cathey2018}, where it was utilized in measurements of the full 6D distribution.

Figure \ref{fig:optics} shows the transverse optics used during measurements.
Energy selection is made after a $90^{\circ}$ dipole, located around $s=3$ meters in Figure \ref{fig:optics}. 
Upstream of the dipole, two vertical slits select thin slices in $x$ and $x'$. 
This ensures a very narrow beam enters the dipole, such that the horizontal spread at dipole exit is created mainly by the beam energy spread.
A third vertical slit downstream of the dipole blocks all but a thin slice in the energy distribution. 
All slits are the same width. The width of the third slit is measured to be $0.17 \pm 0.02$ mm. 
The energy selected by the third slit is a function of the dipole current and the location of the upstream slits. This function is calculated via matrix transformation.

After energy is selected, the beamlet travels through an additional two quadrupoles to the bunch shape monitor (BSM).
The BSM is comprised of a $200\ \mu$m-wide horizontal wire that intersects the beam. 
This wire emits secondary electrons, which are collected and focused onto a microchannel plate. 
Between the wire and plate, an rf deflecting field streaks the beam so that vertical position at the plate corresponds to time-of-arrival.
The microchannel plate amplifies the electron signal which is then imaged via a phosphor screen and camera.  
The signal at the BSM camera is the partial phase distribution $\hat f(\phi)|_{\tilde{x},\tilde{x}',\tilde{y}_2}$ , representing the fraction of beam selected by the vertical slits and BSM wire. 
Thanks to the sensitivity of the BSM screen and the high bit depth of the BSM camera, a signal-to-noise ratio of $10^{3.22}$ was achieved. % inverse of $6 \times 10^{-4}$
The BSM concept is explained in more detail in \cite{Zhang2019,Aleksandrov2013,Feschenko2008}.

\begin{figure}[tbh]
	\centering
	\includegraphics[width=0.4\textwidth]{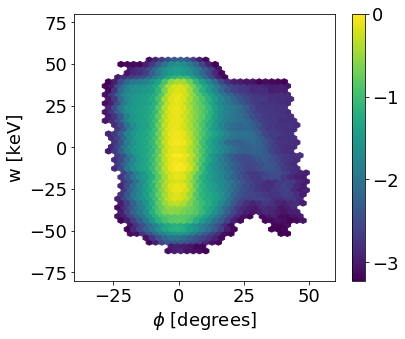}
	\caption{\label{fig:phi-w-center-meas} Measured phase space partial projection near the beam center, $\hat f\left(\phi,w\right)|_{\tilde{x},\tilde{x}',\tilde{y}_2 \sim 0}$. A threshold is applied at $6 \times 10^{-4}$ of the peak density.}
\end{figure}

%%%%%%%% show example measurement
By varying the energy selected by the third slit, the phase space $\hat f(\phi,w)$ can be reconstructed.
The central slice emittance $\hat f(\phi,w)$, which includes the peak density value, is shown in Figure \ref{fig:phi-w-center-meas}. 
In this measurement, phase space is sampled with stepsize $\Delta w = 0.3$ keV.
The phase space ellipse is  upright in the shear-corrected frame in which the linear $\phi-w$ correlation is subtracted.

\subsection{Dimensionality of measurement}

Due to selection by upstream slits and the BSM wire, the measured $\hat f(\phi,w)$ phase space is a partial projection based on only a fraction of the total phase space volume. 
As shown above in Figure \ref{fig:slice-width}, for the expected distribution the core emittance could be as much as $10\%$ lower than the full rms emittance.
The partial projection measured with BTF longitudinal emittance apparatus is:

\begin{equation*}
\hat f(\phi,w) = \int dy_1 f(x,x',y_1,y_2,\phi,w) |_{\tilde{x},\tilde{x}',\tilde{y}_2}
\label{eq:fhat}
\end{equation*}

\noindent where $\tilde{x}=x_0\pm\Delta x$, $\tilde{x}'=x'_{0}\pm\Delta x'$, $\tilde{y}_2=y_{2,0}\pm\Delta y_2$.

Notice that the vertical coordinates are in a frame $y_1,y_2$ rather than the standard $y,y'$. 
At the BSM location, the vertical slice that is selected is $\tilde{y} = y_{wire} \pm \Delta y$. 
However, the horizontal coordinates are referenced to the location of the first vertical slit, which is upstream of the BSM wire by 2.2 meters, four quadrupoles and one $90^{\circ}$ dipole.
The slice made by the BSM wire is rotated in the $y,y'$ phase space at the reference plane.

\begin{figure}
	\centering
	\includegraphics[width=0.4\textwidth]{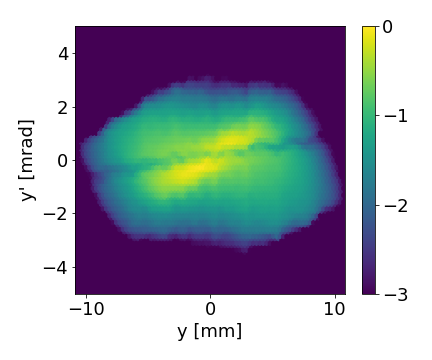}
	\caption{\label{fig:yyp} Vertical phase space at the first slit, showing the BSM wire bisecting the beam core. The intensity scale is logarithmic.}
\end{figure}

Figure \ref{fig:yyp} shows the vertical phase space of the beam at the first slit location, measured using a slit-scan approach.
The shadow of the BSM wire is visible.

%% [duration of scan]
Because the longitudinal emittance apparatus images a 3-fold slice in the six-dimensional phase space, a five-dimensional scan is required to measure the emittance of the ``full" beam.
However,  a five-dimensional scan with high dynamic range and reasonable resolution would have a very long duration.
The measurements have an effective repetition rate of approximately 2.5 Hz, which is half the beam repetition rate. %This extra time cost is due mainly to the latency in actuating the slits. 
At each point in the camera image $f(\phi)$ is averaged for 20 shots in order to improve dynamic range. 
Sampling the phase space on a grid of size 14 x 14 x 14 x 40 in $\left (y_2, x, x', w \right)$ would take an estimated 9 days of continuous measurement.
Instead, this paper takes the approach of conducting four-dimensional scans, iterating over coordinates $\left( x, x', w \right)$. 
A 4D scan requires approximately 16 hours. 
The dependence on the fifth coordinate, $y_2$, is explored by repeating the 4D scan over a range of BSM wire positions.

%#############################################################################
\subsection{Accounting for point-spread increase to measured phase} \label{sec:psf-correction}

\begin{table}[t!]
\centering
\caption{\label{tab:errors} Values for rms point-spread function and $1 \sigma$ errorbars}
\begin{ruledtabular}
\begin{tabular}{lll @{\extracolsep{\fill}} }
Quantity & $w$ [keV] & $\phi$ [$^{\circ}$]\Tstrut\Bstrut\\ \hline
uncertainty & $0.4$ & $0.6$ \Tstrut \\
point-spread & $0.6$ &  $3.3$ \\
\end{tabular}
\end{ruledtabular}
\end{table}

\begin{figure}[t!]
	\centering
	\includegraphics[width=0.4\textwidth]{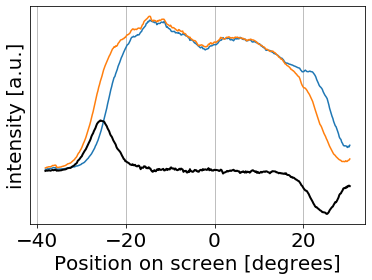}
	\caption{\label{fig:virtual-slit} Illustration of virtual slit concept. Two phase profiles (thin curves) are measured with the BSM camera for two dipole magnet settings separated by 0.05 A. 
	The profiles plotted are obtained with a much wider slit (1 mm) than the standard 0.2 mm slit used for emittance measurements, and the phase profile nearly fills the camera frame. 
	The heavy black line is the differential profile, which recovers two narrower profiles corresponding with the the two edges of the wide profile.}
\end{figure}

\begin{figure}[tbh]
\centering
	\centering
	\includegraphics[width=0.45\textwidth]{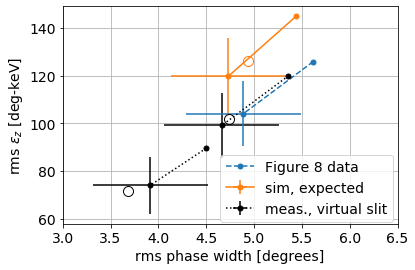}
\caption{\label{fig:analytic-sim-error} 
Comparison of analytic error estimate with actual errors for simulated and measured emittance measurements. 
The lines connect the raw rms values of the reconstructed phase space (higher values) and the value after applying correction, including $1\sigma$ errorbars.
The open circles show the ``true" emittance and phase width, as determined from the virtual slit measurement and simulated emittance reconstruction.
The calculated correction for the phase space shown in Figure \ref{fig:phi-w-center-meas} is included, for which the ``true" values are not known. 
The separation between emittance values at the same phase width is due to the difference in energy widths, which vary between 20 keV and 26 keV. }
\end{figure}

As noted in Section \ref{sec:expected}, space charge defocusing causes a narrowing in phase downstream of the RFQ. 
This brings the phase width close to the phase resolution of the measurement. 
The phase resolution is not limited by the resolution of the BSM, but by the point spread function originating from the finite slit widths. 

Point spread is a systematic, asymmetric error that acts to inflate the measured rms values. 
While the total point-spread is the combination of the three vertical slits, BSM wire and internal BSM electron focusing, the dominant contribution is the width of the third (energy) slit, which affects measurement of both phase and energy. 
Table \ref{tab:errors} summarizies the rms point spread widths, as well as systematic uncertainty originating primarily from uncertainty in calibration curves.
As seen in Table \ref{tab:errors}, the rms phase point-spread is much larger than the uncertainty, and comparable to the expected rms width $5.6^{\circ}$.
The origin and calculation of errors are discussed in more detail in Appendix \ref{ap:errors}.

The majority of the $3.3^{\circ}$ point-spread is due to the large $\phi-w$ correlation at the BSM plane.
For the selected energy slice, the measured phase profile will be wider than the shear-corrected profile that the apparatus is intended to measure.
In comparison, the effect of the energy point-spread is negligible, as the estimated $0.6$ keV rms energy spread selected by the 0.2 mm slit is much smaller than the 23 keV expected width. 
While future improvement may be possible through installation of a narrower slit, given the present limitations of the apparatus it is necessary to estimate the correction to the point spread on the measured rms phase width.

For a Gaussian distribution and point-spread function, the inflated rms values can be corrected through subtraction in quadrature, e.g. $\left< \phi^2 \right > = \left< \phi_{meas.}^2 \right >- \left< \phi_{p.s.f.}^2 \right >$. 
However, both the expected distribution described in Section \ref{sec:expected} and the point-spread function have significantly non-Gaussian features. 
Therefore, in this analysis correction to the rms values is estimated on the basis of simulated and measured recovery of the ``true" phase width, which suggest a much smaller correction than estimated through propagation of Gaussian errors.

% -- dependence on RFQ amplitude
\begin{figure*}[t]
\centering
\begin{subfigure}[t]{0.49\textwidth}
	\centering
	\includegraphics[width=0.95\textwidth]{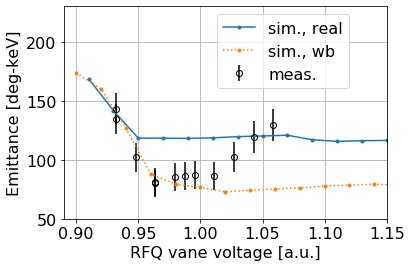}
	\caption{\label{fig:rfq-scan-a} Measured slice emittance compared to fully-projected simulation emittance. Both calculations apply 1\% threshold.}
\end{subfigure}%
\begin{subfigure}[t]{0.49\textwidth}
	\centering
	\includegraphics[width=\textwidth]{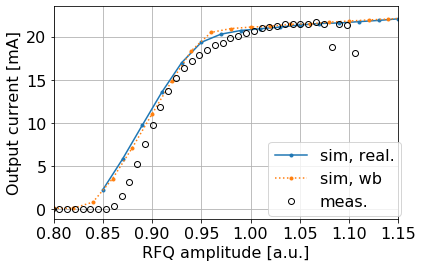}
	\caption{\label{fig:rfq-scan-b} Measured vs. simulated transmitted current. As the RFQ has degraded transmission, and the ion source was producing less than 50 mA at the time of measurement, the simulated current is rescaled to saturate at the same output current.}
\end{subfigure}
\caption{\label{fig:rfq-scan} Dependence of transmission and emittance on RFQ vane voltage.  Results are compared with PARMTEQ simulation of the realistic LEBT distribution as well as an rms-equivalent waterbag, both at 50 mA input. }
\end{figure*}

% #### simulated measurement
In simulation, the point spread error is esimated by propagating the expected distribution through a PyORBIT model of the longitudinal phase space apparatus, including all three slits as shown in Figure \ref{fig:optics}. Details of the approach are included in Appendix \ref{ap:simulate-psf}. 
In measurement, it is possible to obtain sub-slit phase resolution through application of a novel \textit{virtual slit} method.

The virtual slit method requires collecting two phase profiles separated by a differential step in energy slit position and subtracting one from the other. 
The difference waveform includes peak and an anti-peak aligned with the leading and trailing edges of the phase profile, as illustrated in Figure \ref{fig:virtual-slit}.
The difference profiles correspond to the phase profile of a beamlet selected by a virtual slit of width equal to the step size. 
The technique is analagous to the use of scrapers in beam profile measurements, in which transmission is measured as a function of scraper position and differentiated to recover the spatial profile.
As the technique doubles data collection time and reduces dynamic range, it is not applied to the measurements reported in Section \ref{sec:measurements}. 
More details on the implementation of the virtual slit technique are discussed in Appendix \ref{ap:virtual-slit}.

Comparison of the ``true" to ``measured" rms values in both simulated reconstruction and virtual slit measurement allows determination of an appropriate correction factor. In this case, a multiplicative correction to the rms phase, energy and emittance reduces the systematic, point-spread error to well within the uncertainty interval.
As expected, the point-spread function has a relatively small effect on the near-flat-topped energy distribution: the ``true" rms energy width was roughly 95\% the raw ``measured" width in both simulation and experiment. 
The correction to phase width is larger, as expected. Additionally, the required correction has a threshold dependence; as more tails are included in the rms calculation, the relative point-spread error is smaller. 
At 1\% threshold, a correction factor of 87\% minimizes the residual error in simulation and experiment.  At 10\% threshold, the corrected value decreases to 83\% of the raw width. 
There is not enough dynamic range in simulation (limited by particle count) and measurement (limited by virtual slit method) to recommend a correction at the 0.1\% threshold. For the analysis here, the same 87\% correction factor is applied to the 0.1\% threshold values.

Figure \ref{fig:analytic-sim-error} illustrates the magnitude of the rms phase and emittance correction against the ``true" error. 
The one-sigma uncertainty interval is plotted as well; the uncertainty on emittance is estimated by Gaussian error propagation of Table \ref{tab:errors} values under the assumption $\epsilon_z \approx \Delta \phi \Delta w$. (This is valid in the upright, shear-corrected frame as apparent from rms values in Table \ref{tab:sim-phi-w}.)
Applying the estimated corrections to the measured slice emittance shown in Figure \ref{fig:phi-w-center-meas}, the rms values are:
\begin{itemize}
  \item[] rms $\epsilon_{\phi} = 126 - 22 \pm 14$ deg-keV, 
  \item[] rms $\phi = 5.6^{\circ} -0.7^{\circ} \pm 0.6^{\circ}$ and 
  \item[] rms $w = 22.4- 1.1 \pm 0.4$ keV. 
\end{itemize} 
\noindent Here and in all further reported emittance values, the point-spread correction is explicitly stated.

%#############################################################################

% full page figure
\begin{figure*}[p]
\centering 
\begin{subfigure}[b]{\textwidth}	
	\includegraphics[width=0.8\textwidth]{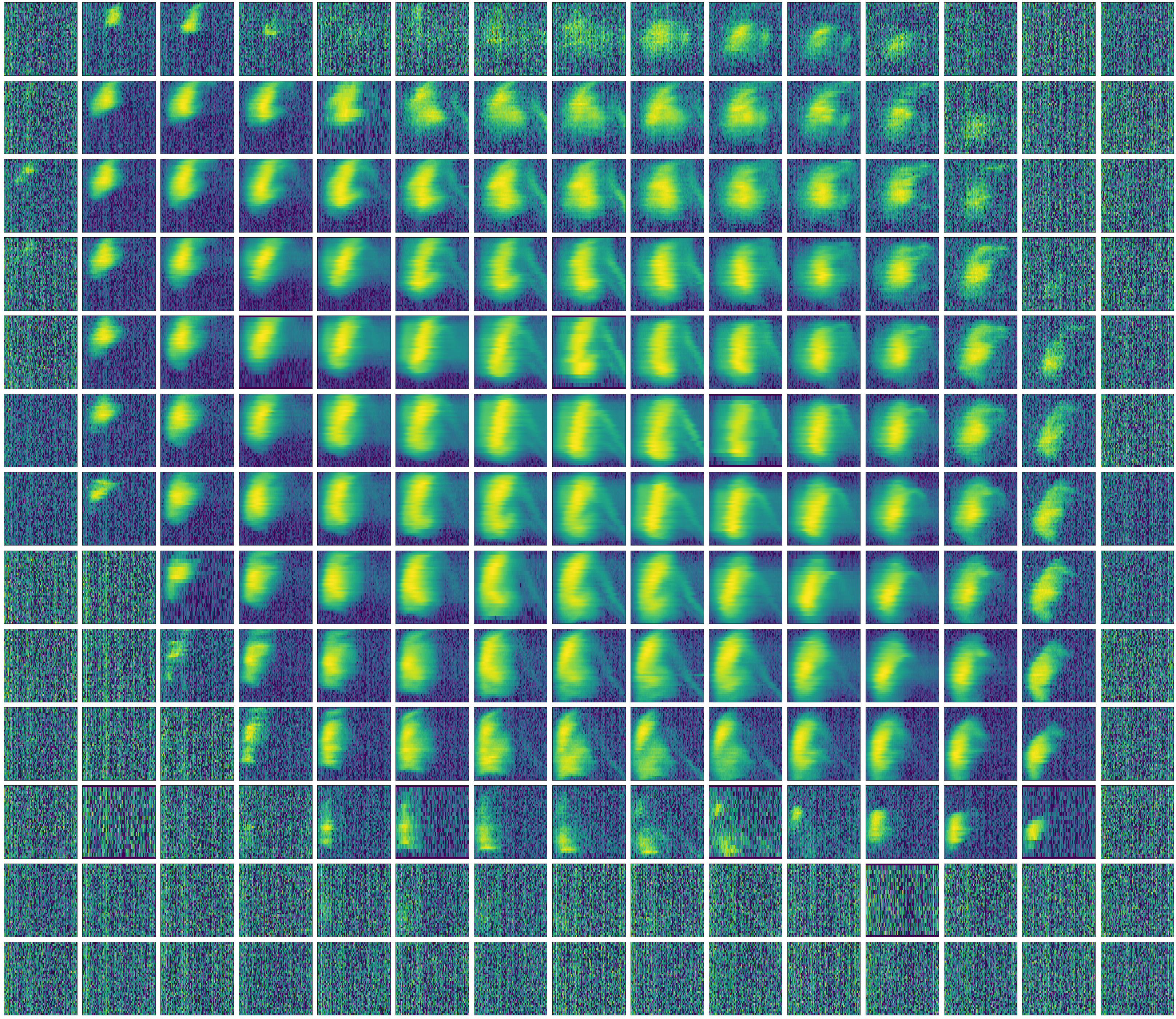}
	\caption{\label{fig:4Dpanels} Minimally-processed data from the 4D scan of $\hat f\left(x,x',\phi, I \right)|_{\tilde{y}_2}$. Each frame shows a partial projection $\hat f(\phi,I)|_{\tilde{x}_1,\tilde{x}'_2}$, with vertical axis $I$ and horizontal axis $\phi$. The axis limits are held fixed for all subplots, but the color scale is not. Each sub-frame corresponds to a different location in $x,x'$. Color is signal strength in logarithmic scale. Data has been cleaned of spurious signals and averaged.}
\end{subfigure}%
\vspace{1ex}
\centering
\begin{subfigure}[t]{0.5\textwidth}
	\centering
	\includegraphics[width=0.9\textwidth]{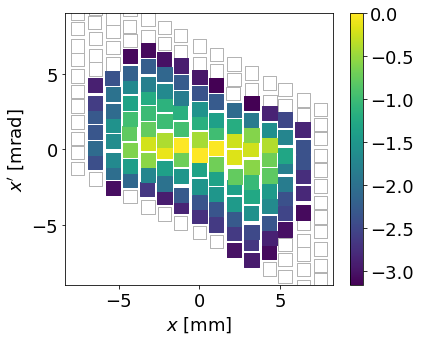}
	\caption{\label{fig:xxp} Image of the 4D scan along the $x, x'$ axes, which are referenced to the location of the first vertical slit. The color of each point is the integrated signal in the $\phi,w$ dimensions. Points with no signal above $10^{-3.22}$ threshold are not filled. The intensity scale is logarithmic.}
\end{subfigure}\hfill%
\begin{subfigure}[t]{0.5\textwidth}
	\centering
	\includegraphics[width=0.9\textwidth]{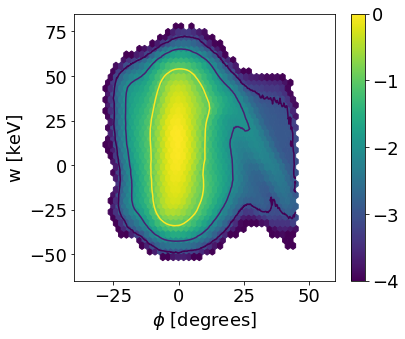}
	\caption{\label{fig:w-phi-int} Longitudinal phase space integrated over horizontal coordinates, $\int dx dx' \hat f\left(\phi,w\right)|_{\tilde{y}_2}$. 
	Contour levels are shown for the three threshold cuts at 0.1\%, 1\% and 10\%}
\end{subfigure}\hfill
\caption{\label{fig:4D} Data from a 4D scan with BSM wire positioned for peak signal strength near the core of the beam.  }
\end{figure*}

%#############################################################################
\section{Measurements} \label{sec:measurements}

Measurements of the longitudinal emittance, which as described above represents a high-dimensional slice in the transverse phase spaces, are repeated many times to map the dependence on several parameters.
First, emittance is measured over a range of RFQ voltages,  which is a free parameter that may be set to obtain minimum output emittance. 
Second, a four-dimensional scan is used to map dependence of the emittance on the transverse dimensions. 
This is then integrated to reconstruct the lower dimensional partial projection $\hat f(\phi,w)|_{\tilde{y}_2}$.
Finally, the 4D scan is repeated for several BSM wire locations, to measure dependence on the coordinate $\tilde{y}_2$.
Measurements of the longitudinal emittance are done at nominally 20-25 mA average current out of the RFQ.

\subsection{Dependence on RFQ amplitude}

The longitudinal phase space is determined by the RFQ parameters. 
Particularly, the RFQ vane voltage may be tuned to produce the optimal (minimal) output emittance. 
For each voltage amplitude, the slice emittance $\hat f(\phi,w)|_{\tilde{x},\tilde{x}',\tilde{y}_2}$ is measured for a fixed $\tilde{x},\tilde{x}',\tilde{y}_2$ slice. 
The slice center in each dimension is chosen to be near the peak density.

Figure \ref{fig:rfq-scan} shows the result of varying RFQ amplitude on longitudinal slice emittance as well as the transmitted current, measured on a Faraday cup positioned after two $90^{\circ}$ dipoles. 
The raw emittance values are corrected and uncertainty assigned according to the correction described in Section \ref{sec:psf-correction}. 
Simulated values of the full emittance at the RFQ exit are included for comparison, for both the realistic LEBT and rms-equivalent waterbag initial distributions.
For these simulations, 40,000 macroparticles are used.

While the simulated vane voltage is applied exactly, the test-stand voltage is not measured. 
Instead, the constant of proportionality between setpoint and vane voltage is chosen for maximum overlap with the simulated curves.
Relative vane voltage is with respect to design value of 83 kV.
The minimum measured emittance occurs at relative amplitude 0.96, corresponding to simulation amplitude 80 kV. %0.610
The RFQ amplitude is set to relative amplitude 0.96 for the all measured results reported here.
The simulated voltage is 83 kV unless otherwise indicated.

The measured 1\% emittance is significantly lower than expected for most voltages, particularly near the setpoint with lowest emittance. 
The predictions of waterbag-seeded distributions agree well with measurement for relative voltages near and below 1.
However, the increase in measured emittance at high RFQ voltage is not reproduced in either simulation.
There is agreement in the sharp emittance increase at low voltages, that coincides with formation of a low-energy tail.

\subsection{Integrated 4D emittance}
As described above, the technique for measuring longitudinal emittance requires making three slices of the 6D distribution in the transverse dimensions.
Therefore, the measured longitudinal phase space represents a three-way slice in phase space, $\hat f(\phi,w)|_{\tilde{x},\tilde{x}',\tilde{y}_2}$.
In order to reconstruct the integrated emittance $\hat f(\phi,w)|_{\tilde{y}_2}$, a 4D scan over variables $\left( x,x',\phi,w \right)$ is performed.

The 4D scan is programmed as a nested loop of the three actuators that select the three dimensions $\left( x,x',w \right)$. These are the first two vertical slits and the dipole current: $\left( x_1,x_2,I \right)$. 
The slice $\tilde{y}_2$ was chosen to give the peak signal strength at the BSM, which corresponds to the BSM wire bisecting the core of the beam. 
Figure \ref{fig:xxp} illustrates the resolution of the 4D scan in transverse phase space by plotting the partial projection $\hat f(x,x')|_{\tilde{y}_2}$. 
The minimally-processed 4D scan data  is shown in Figure \ref{fig:4D}.  
Each sub-plot in Figure \ref{fig:4Dpanels} is the phase space $\hat f(\phi,w)|_{\tilde{x},\tilde{x}',\tilde{y}_2}$ for a point in $x,x',y_2$ space, corresponding with the scatter points in Figure \ref{fig:xxp}. 

Figure \ref{fig:w-phi-int} shows the same data integrated over $x$ and $x'$ to construct the 1D partial projection, $\hat f(\phi,w)|_{\tilde{y}_2}$. 
In addition to integration, significant processing of the data has been done, including thresholding, correcting for variation in microchannel plate response and slow drifts in phase and RFQ output current. 
The output current over the 15.3 hour scan duration was on average $20.5 \pm 0.1$ mA. 
The emittance of the 1D slice with 1\% threshold is $121 - 20 \pm 12$ deg-keV. 
This can be compared to the emittance for the central frame only, $126 - 22 \pm 14$ deg-keV. 
As expected from realistic simulations (Figure \ref{fig:slice-width}, Table \ref{tab:sim-phi-w}), the emittance of a 3D core slice is very close to the emittance of the lower-dimensional 1D slice.

\subsection{Dependence on vertical slice}
 
\begin{figure}[tb!]
\includegraphics[width=0.45 \textwidth]{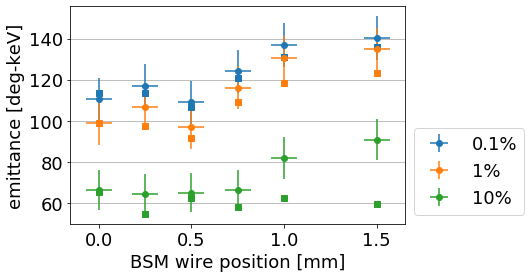}
\caption{\label{fig:bsmwire} Dependence of longitudinal emittance on position of BSM wire. The rms emittance is calculated for three threshold levels for both the 1D slice $\hat f(\phi,w)|_{\tilde{y}_2}$ (shown with errorbars) as well as 3D slice $\hat f(\phi,w)|_{\tilde{x},\tilde{x}',\tilde{y}_2}$ (square points without errorbars) at fixed location $\tilde{x},\tilde{x}'\sim 0$, $y_2 \propto$ BSM wire. }
\end{figure}

The integrated 4D emitttance shown in Figure \ref{fig:4D} is still a partial projection, due to the intersection of the BSM wire with the vertical phase space. 
Dependence on the BSM wire location is measured by repeating the 4D scan procedure at different wire positions.
Figure \ref{fig:bsmwire} shows the resulting rms slice emittances for $\hat f(\phi,w)|_{\tilde{y}_2}$ and $\hat f(\phi,w)|_{\tilde{x},\tilde{x}',\tilde{y}_2}$ versus wire position.
BSM wire position is reported in terms of distance from beam center at the plane of the wire. 
The center is determined to be the BSM wire position with the highest recorded signal intensity, with a precision $\pm 0.25$ mm. 

There is a clear trend of lower emittances in the 1D $y_2$ slice near the core compared to edge slices. 
For the 0.1\% and 1\% emittances, the emittance of the high-dimensional slice $\hat f(\phi,w)|_{\tilde{x},\tilde{x}',\tilde{y}_2}$ is within error-bars of the single-slice $\hat f(\phi,w)|_{\tilde{y}_2}$ emittance, reinforcing the observation that the core slice emittance has low dependence on slice dimensionality.
This breaks down at the 10\% threshold, where the high-dimensional 3D slice emittance is noticeably lower than the 1D slice and has a flat dependence on transverse position.

%\vfill\eject % column break
%#######################################################################################################
\section{Comparson of measurement to simulation} \label{sec:comparison}

% -- measured values with phase wiggle correction
% -- 6/8/20 updayed with 5M bunch; 60k particles in x,x' bunch; 1840 particles in x,x',y bunch; slit width 2x experiment. (x,x' rms are the same with narrow slit width)
\begin{table}[t]
\centering
\caption{\label{tab:phi-w-compare-partial} Comparison of simulated  (expected values) and measured emittances for partially projected phase space $\hat f(\phi,w)|_{\tilde{x},\tilde{x}',\tilde{y}_2=0}$.
Expected distribution values have slice width twice that of physical slit and wire width, for improved particle statistics.
Comparison is not made for $0.1\%$ threshold due to low number of particles in 3D slice. }
\begin{ruledtabular}
\begin{tabular}{llccr}
Threshold & Quantity & Measured & Expected \Tstrut\Bstrut\\
\hline  
0.1\%		& $\epsilon_z$ [deg.-keV] 		&$147 - 25 \pm  13$ 	& -- &\Tstrut \\ 	
1\% 		& $\epsilon_z$ [deg.-keV] 		&$126 - 21\pm 13$ 		& 131  &\\  
10\% 		& $\epsilon_z$ [deg.-keV] 		&$81-17 \pm 11$		& 96 &\\	
1\% 		& rms $\phi$ [deg.] 		&$5.6- 0.7 \pm  0.6 $  	& 5.0 &\\
1\% 		& rms $w$ [keV] 			&$22.4- 1.1 \pm  0.4$ 	& 26.0 &\\
\end{tabular}
\end{ruledtabular}
\end{table}

% -- measured values without wiggle correction
% -- 6/8/20 updayed with 5M bunch; 70k particles in y bunch; slit width= experiment; bins=31,31
\begin{table}[t]
\centering
\caption{\label{tab:phi-w-compare-full} Comparison of simulated  (expected values) and measured emittances for partially projected phase space $\hat f(\phi,w)|_{\tilde{y}_2=0}$.
The slice applied to the simulated (expected) distribution is comparable to BSM wire width.}
\begin{ruledtabular}
\begin{tabular}{llccr}
Threshold & Quantity & Measured  & Expected \Tstrut\Bstrut\\%& Expected, full
\hline  
0.1\%		& $\epsilon_z$ [deg.-keV] 		&$133 - 23 \pm  12$ 	& 122& \Tstrut \\%&130	
1\% 		& $\epsilon_z$ [deg.-keV] 		&$119- 20 \pm 12$ 		& 114& \\%&122	
10\% 		& $\epsilon_z$ [deg.-keV] 		&$86 - 18\pm 11$		& 86&\\%&88	
1\% 		& rms $\phi$ [deg.] 		&$5.6- 0.8 \pm  0.6 $  	& 5.1&\\%&5.4	
1\% 		& rms $w$ [keV] 			&$21.0- 1.0 \pm  0.4$ 	& 22.7&\\%&22.8	
\end{tabular}
\end{ruledtabular}
\end{table}

% -- thresholded emittances; 
\begin{figure}
\centering
\includegraphics[width=0.42\textwidth]{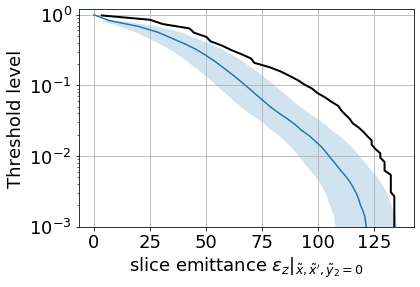}
\caption{\label{fig:ez-threshold} 
Dependence of measured (blue curve, with uncertainty interval) and simulated slice emittances (black curve) for slice $\tilde{x},\tilde{x}',\tilde{y}_2=0$. 
The correction to point-spread error is applied to measured values.}
\end{figure}

\begin{figure*}[htb]
%\begin{subfigure}[t]{0.5\textwidth}
	\includegraphics[width=\textwidth]{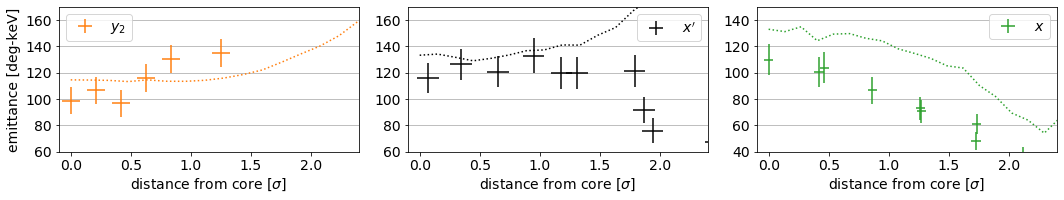}
	\caption{\label{fig:1dslices} 
	Dependence of 1\% rms emittance on position of slices in the transverse coordinates, compared against simulation of the expected distribution (dotted lines) 
	The left-most plot shows emittances for a 1D slice $\tilde{y}_2$. The measured points are the same as shown in Figure \ref{fig:bsmwire}.
	The remaining two plots show emittance for the 3D slice $\tilde{x},\tilde{x}',\tilde{y}_2$.
	In the middle, the center of slice $\tilde{x}'$ is varied while keeping $\tilde{x},\tilde{y}_2=0$.
	On the right, the center of slice $\tilde{x}$ is varied for $\tilde{x}',\tilde{y}_2=0$.
	Distance from core is normalized to rms beam width, to account for difference in simulated and measured transverse beam size.}
\end{figure*}

\begin{figure}[htb]
	\includegraphics[width=0.5\textwidth]{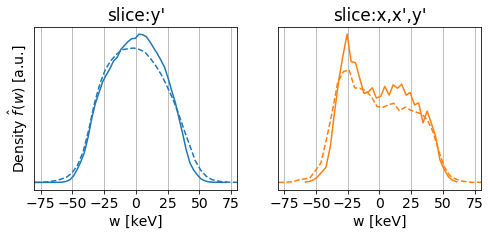}
	\caption{\label{fig:energy-profile-compare} Comparison of partially projected energy profile for measurement (solid lines) and simulation (dashed lines) for 1D (left plot) and 3D (right plot) slices. Here, $\vec{y}_2 \approx \vec{y}'$ is approximated. } 
\end{figure}

The purpose of this study was to evaluate the degree of confidence that can be placed on predictions from ``end-to-end" acccelerator models that include RFQ dynamics. 
The method is to compare a measured beam distribution with output from RFQ simulations, using both rms values and internal structure as metrics. 
In general, the measured rms emittance was 20-30\% lower than expected, a discrepancy that exceeds the $1\sigma$ uncertainty. 
This was illustrated in the previous sections in the comparison of RFQ voltage dependence (Figure \ref{fig:rfq-scan}).
In this case, RFQ simulation seeded with a ``less realistic" waterbag distribution was a better predictor of the output emittance. % and slice dependence (Figure \ref{fig:bsmwire}). 
More detailed comparison to the expected distribution, shown in Tables \ref{tab:phi-w-compare-partial} and \ref{tab:phi-w-compare-full}, shows that this discrepancy persists at all threshold levels. 
This is visualized further in Figure \ref{fig:ez-threshold}.

Although the rms emittances were not reproduced, there was qualitative agreement of the internal, high-dimensional bunch structure. 
In measurement, the emittance of a core slice in $y_2$ was shown to be lower than the edge slices by $\sim 30$ deg-keV (shown in Figure \ref{fig:bsmwire}). 
In analysis of the 4D scan data, the same convex dependence can be seen along coordinate $x'$, but not $x$. Along $x$, the emittance monotonically decreases with distance from core.   
The same general behavior is seen in the expected bunch, as plotted in Figure \ref{fig:1dslices}. 
However, despite similar transverse rms parameters between simulation and experiment, the scale of this feature does not agree; while the width of the simulated feature is $\sim 2\sigma$, in measurement it is closer to $0.5 \sigma$, where $\sigma$ is the rms width in transverse coordinate.

As previously discussed, the rms emittance has a weak dependence on dimensionality. 
In addition, the shape of the high dimensional profiles are in qualitative agreement. 
Figure \ref{fig:energy-profile-compare} compares the measured 1D and 3D partial energy projections against simulated distributions. 
The general shape is reproduced, particularly for the 3D slice profile which is lop-sided with a peak on the low energy side.
Fine-tuning of the energy width can be done by adjusting the current in simulation,  as the initial head-tail deceleration is driven by space charge. 
However this effect cannot explain the emittance discrepancy, as this process does not lead to significant emittance growth. 

Finally, one prominent feature not recreated in simulation is the tail trailing the main bunch. 
This feature is very visible, for example in Figures \ref{fig:phi-w-center-meas} and \ref{fig:w-phi-int}. 
The tail is included in the 1\% and 0.1\% emittance calculations, but excluded when a 10\% threshold is applied.
This can be seen in the emittance curve in Figure \ref{fig:ez-threshold}, where a slight knee is visible just under the 10\% threshold level. 
The amplitude of the tail diminishes with RFQ amplitude; it is possible this is an artifact of non-optimal RFQ voltage that may vanish completely at a higher setpoint.

%#############################################################################
\section{Discussion} \label{sec:conclusion}

The question driving this research is: what is the best strategy for defining an initial distribution, particularly when high accuracy for loss-level predictions is desired?
As seen, the output bunch from the RFQ includes significant internal structure. 
The end-to-end simulation approach provides a high degree of information, both in terms of resolution and interplane correlations. 
However, the complexity of RFQ dynamics means that small errors may result in large discrepancy, as seen here with the rms emittance.
A measurement-based approach avoids this drawback, but with the challenge of typically lower resolution and dimensionality.

In this study, the simulated longitudinal rms emittance was sensitive to the initial LEBT distribution, to a degree that exceeded the measurement uncertainty.
Interestingly, the measured emittance was nearest the predictions from an idealized 4D waterbag.
The ``most realistic" PARMTEQ simulation, based on measured LEBT distribution, predicted rms emittances 20-30\% larger than measured.
This is consistent with errors typically seen in RFQ benchmarking (eg, \cite{Lallement2014,Normand2019}), and is an improvement on the 80\% discrepancy found with independent measurements in the SNS MEBT \cite{Shishlo2018}, but is still unsatisfactory.
Despite the discrepancy, the simulation qualitatively reproduced the observed high-dimensional structure.  

The failure to achieve rms-level accuracy is likely due to errors in the simulation parameters, which may be amplified by nonlinear effects in the RFQ.
The most likely source of error is the LEBT distribution, which was created from quite an old measurement. 
A better understanding of this distribution, particularly the variability during operation and between source changes, may allow for better agreement.

Another candidate is the simulated beam current, which operationally was significantly lower than the design value.
The realistic LEBT distribution was measured at 50 mA, and this was used as the input current for PARMTEQ simulations.
However, at the time of measurements, the LEBT current was measured to be near 40 mA.
In addition, the transmission of the BTF RFQ is significantly lower than the design value. During these measurements it was operating around 60\%, compared to $\geq82\%$  seen in simulation. 
Through combination of these effects, the measured current in the MEBT is about 50\% less than the current of the expected bunch. 
While this discrepancy may account for some differences in the core structure, the simulated rms emittance was not seen to depend strongly on the input LEBT current.
In MEBT simulations, the energy-phase aspect ratio increases with space charge but the rms emittance has a flat dependence.

Finally, as mentioned the RFQ vane voltage is not precisely known, and likely contributes a systematic error to the simulated value. 
As discussed in the previous section, a lower-than-optimal operating voltage can account for the presence of tails in the measurement. 
It is also possible operating at a higher RFQ voltage may also result in better agreement in rms emittance, as the measured voltage dependence shows an increasing trend at higher voltages. 
This will be addressed in future studies, through implementation a non-intrusive bremsstrahlung voltage diagnostic \cite{Zhukov2016a}.

Another limitation of this study was the phase resolution, which is too low to accurately measure the beam phase profile. 
The system was originally designed on the basis of the output from RFQ simulations, which as shown in Section \ref{sec:parmteq} has nearly equal aspect ratio. 
However, due to space charge defocusing in the first meter of the MEBT, the longitudinal bunch rapidly elongates in phase space. At the plane of the BSM the phase profile is considerably narrower than at the RFQ exit. 
The experiment resolution can be improved by either decreasing the width of the energy slit or reducing the linear $\phi-w$ correlation at the energy slit location via installation of a rebunching cavity upstream of the BSM. 

As beam halo mitigation becomes a more pressing matter for high-intensity accelerators, the demand for predictive accelerator models will grow.  
As such, the question of generating realistic and representative distributions needs to progress beyond rms equivalence and 2D characterization. 
For planned studies in halo evolution at the SNS BTF, agreement with dynamic range $\geq 10^4$ is sought.
The bunch produced via PARMTEQ simulation does not benchmark at the 10\%-0.1\% threshold level, and therefore is not trusted to deliver good halo predictions.
Ongoing work is focused on generating an initial MEBT bunch on the basis of direct 6D measurement.

Full 6D characterization is still an impractical solution for wide application. 
As such, RFQ simulations will continue to be a powerful tool for generating fully-correlated distributions.
From previous efforts, rescaling bunch coordinates to match measured rms widths may not be enough for predicting downstream evolution \cite{Qiang2002}. 
Looking forwards, the need for improved simulation accuracy will require more sophisticated strategies for generating distributions. 
This will entail reconciling simulated 6D coordinates with both low- and high-dimensional measurements, and extending the metrics for achieving agreement beyond rms equivalency. 

%%%%%%%%%%%%%%%%%%%%%%%%%%%%%%%%%%%%%%%%%%%%%%%%%%%%%%%%%%%%%%%%%%%%%%%%%%%%%%%%%%
\subsection{Acknowledgments}
%%%%%%%%%%%%%%%%%%%%%%%%%%%%%%%%%%%%%%%%%%%%%%%%%%%%%%%%%%%%%%%%%%%%%%%%%%%%%%%%%%
The authors acknowledge the contributions of Brandon Cathey, who not only authored the initial high dimensional beam study \cite{Cathey2018} but also the data collection software used for the 4D characterization reported here.  
The authors are also grateful for the assistance of SNS operations, whose personnel monitored data collection during long study times. 
This manuscript has been authored by UT-Battelle, LLC under Contract No. DE-AC05-00OR22725 with the U.S. Department of Energy. 
This research used resources at the Spallation Neutron Source, a DOE Office of Science User Facility operated by the Oak Ridge National Laboratory.
The United States Government retains and the publisher, by accepting the article for publication, acknowledges that the United States Government retains a non-exclusive, paid-up, irrevocable, world-wide license to publish or reproduce the published form of this manuscript, or allow others to do so, for United States Government purposes. The Department of Energy will provide public access to these results of federally sponsored research in accordance with the DOE Public Access Plan(http://energy.gov/downloads/doe-public-access-plan).

\appendix
%%%%%%%%%%%%%%%%%%%%%%%%%%%%%%%%%%%%%%%%%%%%%%%%%%%%%%%%%%%%%%%%%%%%%%%%%%%%%%%%%%

%#############################################################################
\section{Sources of error} \label{ap:errors}

\begin{figure}[hbt]
\includegraphics[width=0.35 \textwidth]{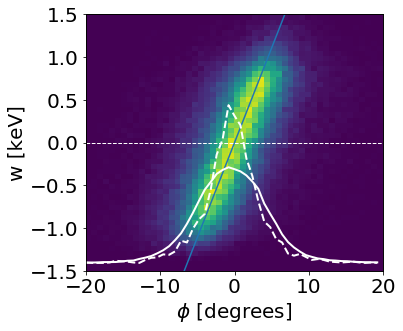}
\caption{\label{fig:beamlet-z-w} $f(\phi,w)$ phase space immediately after selection at the energy slit. The solid white line plots the projected phase distribution $f(\phi)$. The dashed line shows the partial phase distribution $\hat f(\phi)|_{w=0}$ (along the thin horizontal dashed line). This distribution is generated through PyORBIT simulation.}
\end{figure}

% BSM output current: maybe not necessary?
\begin{figure*}
	\includegraphics[width=\textwidth]{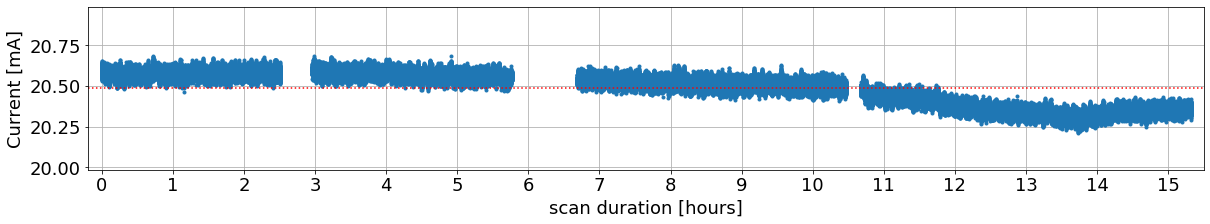}
	\caption{\label{fig:bcm-current} Output at beam current monitor during collection of results shown above. Red dashed line indicates average over scan of 20.48 mA. Pauses are due to routine lock-outs of BTF facility.}
\end{figure*}

There are many potential sources of error. These can be separated into four categories by origin, ordered by effect on the measurement:

\begin{enumerate}
\item \label{it:res} \textbf{Resolution} in energy and phase, determined by the physical width of the three slits as well as the phase resolution of the BSM. 
\item \textbf{Calibration errors}, which are applied in the calculation of energy and phase coordinates. This uncertainty is determined by the variance of a linear least squares fit of the calibration data. % incl absolute energy 
\item \textbf{Model geometry}, including uncertainty in path length and strength of magnetic elements, used in the calculation of energy. This includes uncertainty in machine readbacks such as slit position and dipole current.
\item \label{it:stat} \textbf{Machine variation}, encompassing both slow drifts and jitter. 
\end{enumerate}

\noindent The largest source of error is due to the resolution of the measurement.
The finite slit width create point-spread in both the energy and phase dimensions that result in systematic over-estimation of rms parameters.
While the total point-spread is the combination of the three vertical slits, BSM wire and internal BSM electron focusing, the largest term is the width of the third (energy) slit.
The energy point-spread is relatively small: 0.6 keV compared to the rms width $23$ keV of the expected distribution.%about 4\% of the measured rms energy spread.

However, for the phase coordinate the relative error is much larger. 
The main contribution to phase spread comes from two sources: the finite width of the energy slice and the electron optics in the BSM.
The point-spread of the BSM can be directly measured by disabling the BSM RF deflector and recording the image of the BSM wire. 
The measured rms width of the internal BSM point-spread is $0.9^{\circ}$. 

The energy spread contributes to phase spread at the BSM through time-of-flight. For a collection of particles with rms energy spread 0.6 keV originating at the same phase in the plane of the energy slit, the phase spread at the BSM will be $0.5^{\circ}$. However, there is an additional, larger point spread effect due to the fact that at the energy slit the bunch is already highly correlated. Therefore, the projected phase width is significantly wider than the phase width of a monoenergetic slice. This effect is illustrated in Figure \ref{fig:beamlet-z-w}, which is generated through PyORBIT simulation of the expected distribution to the energy slit.
The 100\% rms phase width is $5.8^{\circ}$, but through projection of the correlated phase space, the apparent width increases to $6.6^{\circ}$. 
Assuming Gaussian phase distribution and point spread function, the rms width of the point-spread from correlation is estimated to be $3.5^{\circ}$.
Adding the three sources of phase spread in quadrature, the total rms point-spread is $3.3^{\circ}$, roughly half the expected width.

\begin{figure*}[!th]
\centering
\begin{subfigure}[t]{0.4\textwidth}
	\centering
	\includegraphics[width=\textwidth]{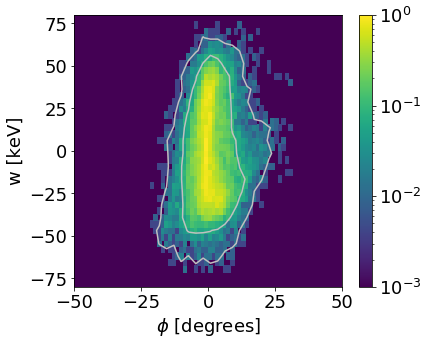}
	\caption{\label{fig:phi-w-center-sim} Simulated phase space at the reference plane (location of first vertical slit).}
\end{subfigure}
\begin{subfigure}[t]{0.4\textwidth}
	\centering
	\includegraphics[width=\textwidth]{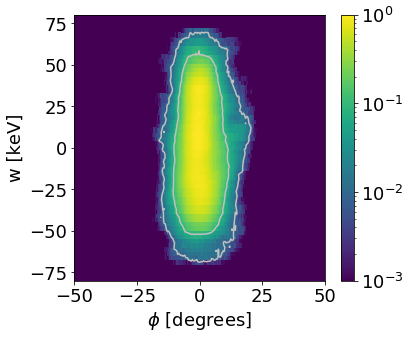}
	\caption{\label{fig:phi-w-center-reconstr} Reconstruction of phase space from simulation of emittance measurement with finite slit widths.}
\end{subfigure}
\caption{\label{fig:phi-w-compare} Comparison between $\hat f\left(\phi,w\right)|_{\tilde{x},\tilde{x}'}$ at the first slit and the reconstructed phase space after simulation of the phase space measurement. The effect of the slit-width point spread on the phase width is apparent. 1\% and 10\% contour lines are drawn.}
\end{figure*}

Calibration errors make the next largest contribution and dominate the calculation of systematic uncertainty. 
This includes calibration of dipole strength ($-1.009 \pm 0.006$ mm/A relative to motion of the beam at the energy slit) and conversion of BSM camera pixels to arrival phase ($0.167 \pm 0.008$ degrees/pixel).
These uncertainties grow linearly with distance from the central phase and nominal dipole current. 
At the rms width of the expected distribution $w=23$ keV, $\delta w = \pm 0.4$ keV and at $\phi = 5.6^{\circ}$, $\delta \phi = \pm 0.6^{\circ}$. 

Finally, uncertainty in the model geometry used to calculate beam energy has a negligible effect on calculated errors. 
Variations in the BTF beam and measured signal also have a negligible contribution.
The effect of jitter (item 5) is reduced through averaging, and the overall statistical uncertainty is low. % could estimate number
Slow variations, including drifts in phase, RFQ output current (Figure \ref{fig:bcm-current}) and BSM micro-channel plate response, are corrected before emittance is calculated.

Assuming that the calibration errors are independent, they can be summed in quadrature to estimate the uncertainty in the measured emittance. The same can be done for the systematic error of the point spread function. 
In the approximation $\epsilon_z \approx \Delta \phi \Delta w$, the error propagates as 

\begin{equation}
\left( \frac{\delta \epsilon_z}{\left<\epsilon_z^2\right>} \right)^2 \sim \left( \frac{\delta \phi}{\left<\phi^2\right>} \right)^2 + \left( \frac{\delta w}{\left<w^2\right>} \right)^2
\end{equation}

%#############################################################################
\section{Simulated estimate of point spread error} \label{ap:simulate-psf}

The analytic error estimate is based on the assumption that the $\hat f(\phi,w)$ distribution and point spread functions are Gaussian, which is very much not true. 
In order to more carefully estimate the systematic error due to finite slit widths, the longitudinal emittance measurement was reproduced with PyORBIT simulation.
The expected distribution from the RFQ is tracked to the location of the BSM wire, with slit apertures applied as in measurement. 
The first two vertical slits are centered at $\tilde{x} =0 \pm 0.1$ mm,  $\tilde{x}' =0  \pm 0.2$ mrad. 
The simulation is repeated for different positions of the third (energy) slit, at a spacing of $\Delta w = 0.25$ keV.
Just as in measurement, the emittance $\hat f(\phi,w)$ is reconstructed by combining these phase distributions.
The selection of vertical phase space at the BSM wire is not included, as should have a negligible effect on the point spread error.

The expected distribution, generated through RFQ simulation with 5,000,000 particles, is resampled back to 5,000,000 to after the first two slit aperatures. This number is chosen to maintain good particle statistics in phase space slices. For a slit positioned at the density peak, only about $2\%$ of particles pass through. 
The re-sampling results in slight artifical growth of the longitudinal emittance at each slit (about 1\% at each). 
This is smaller than the point spread effect.

The reconstructed distribution is compared to the longitudinal distribution in the plane of the first vertical slit in Figure \ref{fig:phi-w-compare}.
No significant space-charge influenced evolution of the longitudinal emittance is expected between this location and the BSM. 
The broadening of the distribution due to finite slit width is apparent, particularly in the phase width.

\begin{figure}[tb]
	\centering
	\includegraphics[width=0.4\textwidth]{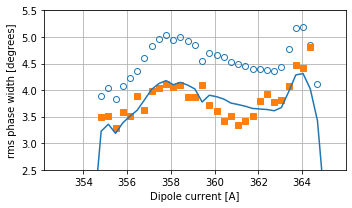}
	\caption{\label{fig:virtual-slit-phi} Comparison of rms phase width at 1\% threshold. Solid orange points are phase width measured with virtual slit technique. Open blue points are result of ``typical" measurement. The solid blue line is the same measurement with 83\% correction factor applied.  }
\end{figure}

\section{Sub-slit resolution with virtual slit} \label{ap:virtual-slit}

% TABLE NUMBERS UPDATED; virtual slit 5/18 (rfq=0.610); numbers from 6/1 are at amplitude 0.640 and emittance is significantly lower.
\begin{table}[htb]
\centering
\caption{\label{tab:virtual-slit} Comparison of virtual slit reconstructed emittance with ``typical" measurement at 1\% threshold.}
\begin{ruledtabular}
\begin{tabular}{llll}
Quantity & typical meas. & virtual slit meas. & uncertainty \Tstrut\Bstrut\\
\hline  
rms $\epsilon_{\phi}$ [deg-keV] 	& 120			& 102 			& 13 \Tstrut \\
 rms $\phi$  				& $5.4^{\circ}$  	& $4.7^{\circ}$  	& $0.6^{\circ}$ \\
rms $w$ [keV] 			& 23.1		& 22.4		& 0.4 \\ 
\end{tabular}
\end{ruledtabular}
\end{table}

As established, the largest error in the emittance measurement is due to the point spread associated with the finite width of the energy slit. 
However, it is possible to obtain a higher resolution that overcomes the physical limitations of the existing apparatus, without the need to manufacture and install narrower slits.
This is done by creating a \textit{virtual slit} from two phase profiles separated by a differential step in slit position.
The step size must be smaller than the physical slit width for enhanced resolution.

% -- reconstruction algorithm

As in the typical emittance measurements, the dipole current is varied rather than actuating the energy slit. 
This is particularly beneficial for the virtual slit measurements, as the dipole current can be set with higher precision ($\pm0.005$ A, equivalent to 0.005 mm response at energy slit) than the slit actuator position ($\pm 0.02$ mm).
A virtual slit spacing of 0.05 A (0.05 mm) was found to be sufficient in both simulation and measurement. 
In application of the virtual slit reconstruction on simulated data, the recovered phase width plateaued for slit separations $\leq 0.07$ was within 5\% of the ``base truth" phase width. 
In measurement, the recovered phase width appears to plateau at 0.05 A separation. 

% [UPDATE AGAIN] numbers have been updated. 
Figure \ref{fig:virtual-slit-phi} compares the measured rms phase widths at 1\% threshold for a range of energies (plotted as dipole current). Phase width is calculated both with and without application of the virtual slit technique. 
The reconstructed phase profiles are significantly more noisy, byt on average (for all profiles measured in this data-set), the reconstructed width is equal to 83\% of the measured width without correction (shown in the figure as a solid line). 

% THESE NUMBERS UP TO DATE.
Table \ref{tab:virtual-slit} compares rms for a ``typical" phase space measurement with the reconstruction using the virtual slit technique.
Note that there is still uncorrected point-spread from the internal BSM optics, but at rms $0.9^{\circ}$ this leads to a much smaller error than from the energy slit.

\bibliography{sns-ez-measurement}

\end{document}